\documentclass[aip,pop,reprint]{revtex4-1}
\usepackage{graphicx}
\usepackage{ulem}
\usepackage{amssymb}
\usepackage{amsmath}
\usepackage{todonotes}
\usepackage{wasysym}
\usepackage{relsize}
\usepackage{bm}
\begin{document}

\title{Verification of shielding effect predictions for large area field emitters }

\author{Rashbihari Rudra }
\affiliation{
Bhabha Atomic Research Centre,
Mumbai 400 085, INDIA}
\affiliation{
Homi Bhabha National Institute, Mumbai 400 094, INDIA}
\author{Debabrata Biswas}
\affiliation{
Bhabha Atomic Research Centre,
Mumbai 400 085, INDIA}
\affiliation{
  Homi Bhabha National Institute, Mumbai 400 094, INDIA}

%\pacs{85.45.-w}{}
%\pacs{03.65.Sq}{}
%\pacs{03.65.Xp}{}
%\pacs{52.59.Sa}{}

\begin{abstract}
  A recent analytical model for large area field emitters, based on the line charge model (LCM), provides a simple
  approximate formula for the field enhancement on  hemiellipsoidal emitter tips in terms of the ratio of
  emitter height and pairwise distance between neighbouring emitters. The formula, verified against the exact solution
  of the linear LCM, was found to be adequate provided the mean separation between emitters is larger than
  half the emitter height. In this paper, we subject the analytical predictions to a more stringent test by
  simulating (i) an infinite regular array and (ii) an isolated cluster of 10 random emitters, using the
  finite element software COMSOL.
  In case of the array, the error in apex field enhancement factor (AFEF) is found to be less than $0.25\%$
  for an infinite array when the
  lattice constant $c \geq 1.5h$, increasing to $2.9\%$ for $c = h$ and $8.1\%$ for $c = 0.75h$. For an isolated
  random cluster of 10 emitters, the error in large AFEF values is found to be small. Thus, the error in net emitted
  current is small for a random cluster compared to a regular infinite array with the same (mean) spacing.
  The line charge model thus provides a reasonable analytical tool for optimizing a large area field emitter.
 
  \end{abstract}

%\email{dbiswas@barc.gov.in}
%\maketitle

%\pacs{85.45.-w}{}
%\pacs{03.65.Sq}{}
%\pacs{03.65.Xp}{}
%\pacs{52.59.Sa}{}

%\date{\today}
%\vskip 0.2 in
%\centerline{\bf Abstract}

%\vskip 0.25 in

%\pacs{85.35.-p, 03.65.Sq, 52.59.Sa}

\maketitle

%]
\newcommand{\be}{\begin{equation}}
\newcommand{\ee}{\end{equation}}
\newcommand{\bea}{\begin{eqnarray}}
\newcommand{\eea}{\end{eqnarray}}
\newcommand{\Tbar}{{\bar{T}}}
\newcommand{\En}{{\cal E}}
\newcommand{\K}{{\cal K}}
\newcommand{\GC}{{\cal \tt G}}
\newcommand{\Lop}{{\cal L}}
\newcommand{\DB}[1]{\marginpar{\footnotesize DB: #1}}
\newcommand{\q}{\vec{q}}
\newcommand{\kt}{\tilde{k}}
\newcommand{\Lopn}{\tilde{\Lop}}
\newcommand{\noi}{\noindent}
\newcommand{\ovn}{\bar{n}}
\newcommand{\ovx}{\bar{x}}
\newcommand{\ovE}{\bar{E}}
\newcommand{\ovV}{\bar{V}}
\newcommand{\ovU}{\bar{U}}
\newcommand{\ovJ}{\bar{J}}
\newcommand{\calE}{{\cal E}}
\newcommand{\ovphi}{\bar{\phi}}
\newcommand{\zt}{\tilde{z}}
\newcommand{\rt}{\tilde{\rho}}
\newcommand{\tth}{\tilde{\theta}}
\newcommand{\nuv}{{\rm v}}
\newcommand{\ck}{{\cal K}}
\newcommand{\cc}{{\cal C}}
\newcommand{\ca}{{\cal A}}
\newcommand{\cb}{{\cal B}}
\newcommand{\cg}{{\cal G}}
\newcommand{\ce}{{\cal E}}
\newcommand{\fn}{{\small {\rm  FN}}}
\newcommand\norm[1]{\left\lVert#1\right\rVert}

%\newpage
%\noindent

%\section{Introduction}
%\label{sec:Introduction}

\section{Introduction}
\label{sec:intro}

Large area field emitters (LAFE) are promising as a high brightness source of cold electrons.
They have been investigated for around four decades, as patterned arrays of pointed emitters
or clusters of nanotubes or nanorods. They find applications in various vacuum devices such as
X-ray tubes, terahertz generators and even in space navigation \cite{spindt68,spindt76,parmee,basu2015,cole2016,whaley2018}.

The electron emission mechanism, at least in metals, is fairly  well understood\cite{FN,murphy,forbes,jensen_ency}
and it is recognized that the enhanced local electric\cite{jensen_ency,forbes2003,db_fef} field at a
curved emitter tip leads to an increase in the quantum
transmission coefficient. Thus, significant currents can be observed even at moderate macroscopic
fields. The problem in bringing together a bunch of curved emitter tips is also well recognized.
The proximity of emitters causes electrostatic shielding and hence the local field enhancement
on emitter tips is not as much compared to isolated emitters\cite{read_bowring,harris15,jap2016,db_rudra}.
Thus, a LAFE must have an optimal
packing density such that the overall current density is maximum.

Knowing the degree of field enhancement for an isolated emitter of given shape, height ($h$) and apex radius
of curvature ($R_a$) is in
itself a formidable task. For a LAFE, shielding makes this all the more complicated. At present, a theory
of LAFE exists for only the simplest of emitters where the shape is hemiellipsoidal and for which the
apex field enhancement factor (AFEF) for an isolated emitter is known analytically\cite{kosmahl,pogorelov}.
The theory predicts the AFEF, $\gamma_a$ of an ($i^{th}$) emitter in a given LAFE environment in terms of distances from all other
emitters. Thus\cite{db_rudra},

\be
\gamma_a^{(i)}  \simeq  \frac{2h/R_a}{\ln\big(4h/R_a\big) - 2 + \alpha_{S_i}}  \label{eq:gamN0}
\ee

\noi
where $\alpha_{S_i} = \sum_{j\ne i} (\lambda_j/\lambda_i) \alpha_{S_{ij}} \simeq \sum_{j\ne i}  \alpha_{S_{ij}}$  and 

\be
\alpha_{S_{ij}}  =  \frac{1}{\delta_{12}}\Big[1 - \sqrt{1 + 4\delta_{ij}^2} \Big] + \ln\Big|\sqrt{1 + 4\delta_{ij}^2} + 2\delta_{ij} \Big| \nonumber
\ee

\noi
with $\delta_{ij} = h/\rho_{ij}$, $\rho_{ij} = [(x_i - x_j)^2 + (y_i - y_j)^2]^{1/2}$ being the distance between on the
cathode plane ($XY$) between the $i^{th}$ and $j^{th}$ emitter. In the above, $\lambda$ is the slope of the line charge density
$\Lambda(z)$ (i.e. $\Lambda(z) = \lambda z$),
obtained by projecting the surface charge density along the emitter axis\cite{jap2016}.

Eq.~(\ref{eq:gamN0}) is approximate since $\alpha_{S_i}$ assumes the charge distribution on
the $i^{th}$ and $j^{th}$ emitter to be identical. This is largely true when they are not too close so that  Eq.~(\ref{eq:gamN0})
serves as a useful first approximation. It can however cause errors for separations smaller than the emitter height.
A second source of errors concerns the nature of the charge distribution on the surface of the hemi-ellipsoid.
An isolated hemiellipsoid with its axis aligned along an external macroscopic field $E_0 \hat{z}$, has a projected
line charge distribution (along the emitter ($Z$) axis) that is linear\cite{jap2016}. However, when such
emitters are close together,
the line charge distribution can develop non-linear components. Since Eq.~(\ref{eq:gamN0}) is based on a linear
model, this may contribute to the error in AFEF when emitters are close together.

Note that Eq.~(\ref{eq:gamN0})
compares well with the exact linear LCM model for an $N$-emitter random LAFE.  For mean spacings larger than $h/2$, the
observed error was $< 6\%$ while for mean spacings larger than $h$, the error is less than $1.5\%$. This
comparison however neglects the non-linearity factor in the charge distribution.
It is thus necessary to subject Eq.~(\ref{eq:gamN0}), obtained using linear LCM, to more stringent tests
such as by comparing its predictions with numerical (finite element) simulations where the projected charge
density does not have constraints of linearity. Such a comparison is also required since LCM predictions are reported to be at
variance with other models/numerical predictions\cite{forbes_2018a}. A reasonable outcome from the present study can pave
the way for a greater reliance on LCMs for analytical investigations of large area field emitters where direct
numerical methods are difficult to implement due to computational constraints.

The paper is organized as follows. We shall first take a look at the computational domain required to
model a large area field emitter using COMSOL with the `anode-at-infinity' and also
study the number of neighbouring emitters required for the convergence of the AFEF using Eq.~(\ref{eq:gamN0}).
Next, we shall compare the predictions of linear LCM (i.e. Eq.~(\ref{eq:gamN0})) with those of COMSOL for an infinite square array.
Finally, we shall also study a bunch of random isolated emitters using COMSOL. Rather than studying just
the error in apex field enhancement
factor of individual emitters in the cluster, we shall also compute the error in net
emitted current so that emitters in close proximity
do not get a disproportionately large weight in deciding the error in LCM prediction. Finally, we shall
discuss the implications of our results in designing large area field emitters.

\section{Domain size for COMSOL and LCM}

The array-at-infinity is an idealization that simplifies the line charge model but is not essential to it. Computationally,
an infinite square array with lattice constant $c$ can be simulated by imposing `zero surface charge density' at $x,y = \pm c/2$.
Thus, $\partial V/\partial (x,y) = 0$ at $x = \pm c/2$ and $y = \pm c/2$. The boundary condition at the anode can be Dirichlet
($V = V_A$, where $V_A$ is the anode potential) or Neumann\cite{agnol} ($\partial V/\partial z = \epsilon_0 E_0$, $E_0$ being the
magnitude of the macroscopic field $-E_0 \hat{z}$)
at $z$ sufficiently far from the emitter tip. A generally accepted guideline is to place the anode at about 5 times the
emitter height in order to impose the Dirichlet boundary condition while for the Neumann condition, the anode can be somewhat
closer. It is important however to test for convergence to the anode-at-infinity result by pushing the anode further away
in both cases.

\begin{figure}[htb]
\vskip -.50cm
\hspace*{-1.00cm}\includegraphics[width=0.6\textwidth]{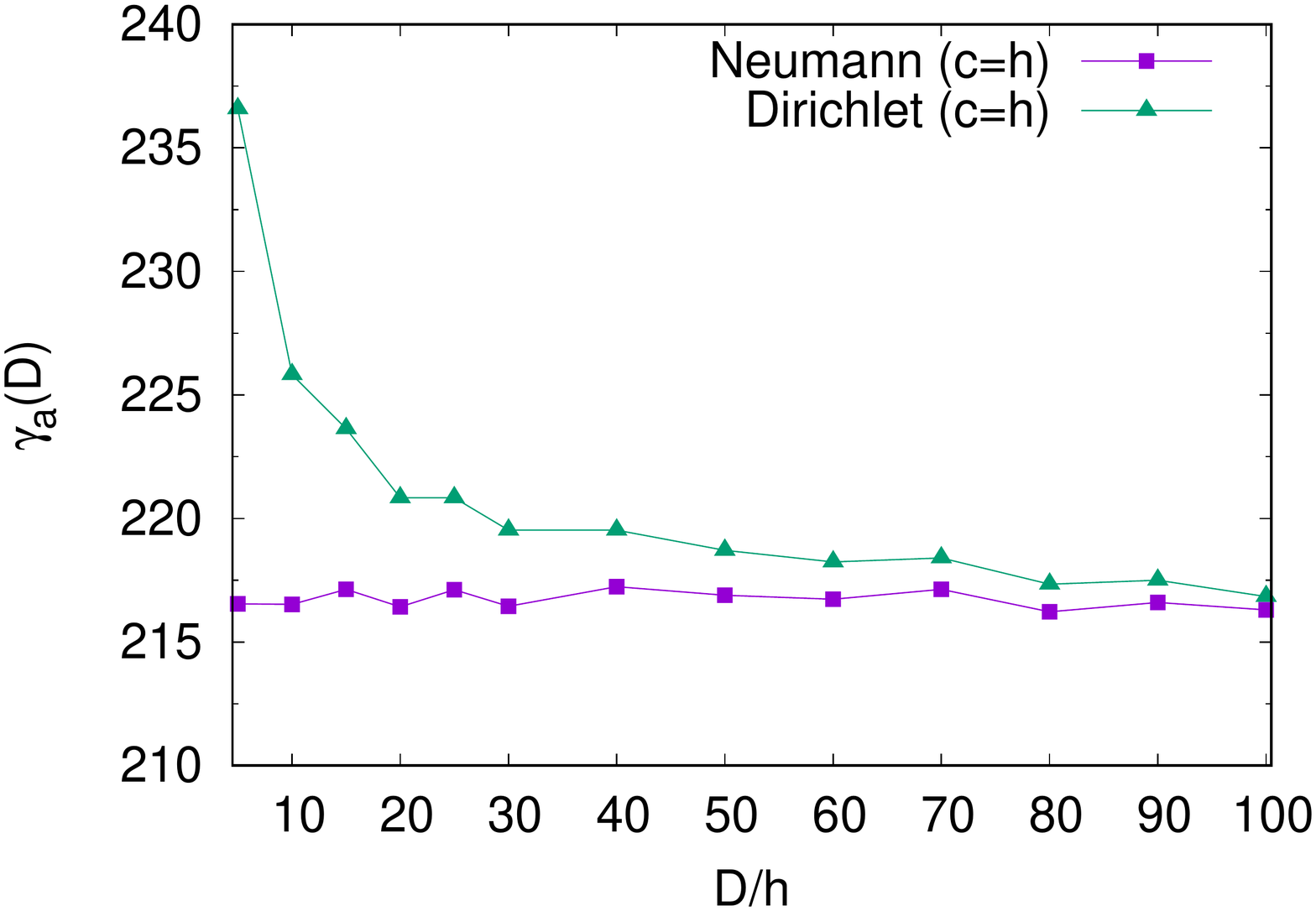}
\vskip -1.0cm
\caption{The apex enhancement factor evaluated using COMSOL with the anode plate having Dirichlet ($V = V_A$) and
  Neumann ($\partial V/\partial z = \epsilon_0 E_0$). The anode-at-infinity is easily achievable with the Neumann
boundary condition. The emitter height $h = 1500\mu$m, lattice constant $c = h$ while $R_a = 1.5\mu$m .}
\label{fig:diri_neu}
\end{figure}

Fig.~\ref{fig:diri_neu} shows a convergence study for an infinite hemiellipsoidal array with $h = 1500~\mu$m, $R_a = 1.5\mu$m
and $c = h$ using COMSOL. The anode-cathode plane distance is increased from $D = 5h$ for both the Dirichlet and
Neumann boundary conditions
and the apex field enhancement factor $\gamma_a$ is plotted against $D$. Clearly, the Neumann boundary condition
achieves the anode-at-infinity condition at a much smaller $D$ value while the Dirichlet condition in this case requires the anode
to be at $D = 100h$. We shall henceforth use the Neumann boundary condition for simulating the anode-at-infinity.

\begin{figure}[htb]
\vskip -1.0cm
\hspace*{-1.0cm}\includegraphics[width=0.60\textwidth]{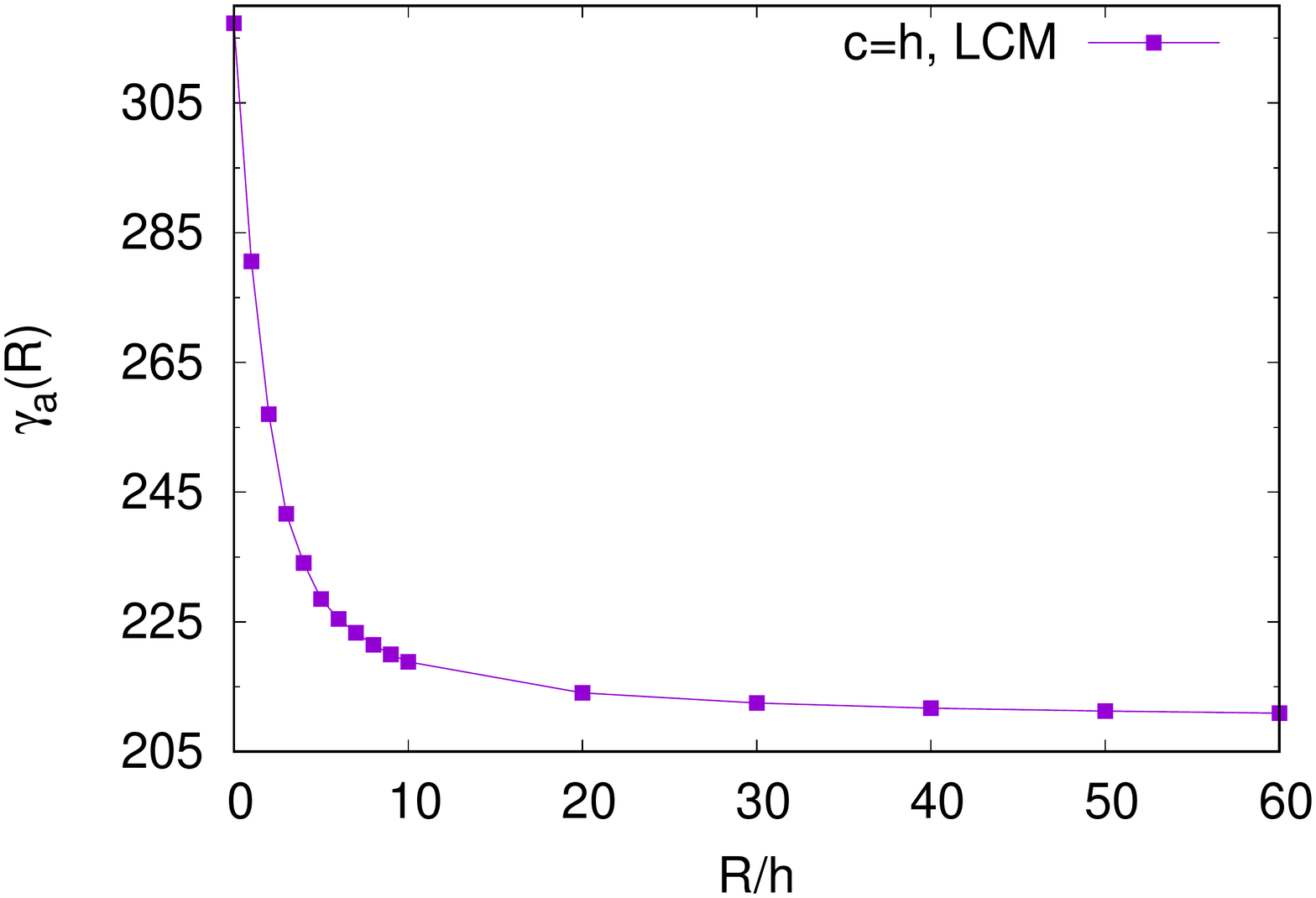}
\vskip -1.0cm
\caption{The apex enhancement factor evaluated using the LCM prediction of Eq.~(\ref{eq:gamN0}).
  All shielding emitters within a circle of radius $R$ are included.
  The emitter height $h = 1500\mu$m, $R_a = 1.5\mu$m while $c = h$.}
  \label{fig:lcm_converge}
\end{figure}

We shall next fix the question of the number of emitters required for convergence of the LCM result, Eq.~(\ref{eq:gamN0}).
Figure \ref{fig:lcm_converge} shows the apex field enhancement factor calculated by including all ($j^{th}$) emitters
in a circle of radius $R$. The $R=0$ limit corresponds to an isolated emitter while at $R/h = 5$ and $c = h$,
the number of emitters is 80. At $R/h = 60$, where approximate convergence is achieved, the number of shielding emitters is 11288.  
In all AFEF calculations henceforth, we shall consider $R/h = 100$ to ensure that convergence in AFEF has been achieved.

\section{The error in LCM prediction}

With the question of domain size settled, we are now in a position to investigate the error in LCM prediction.
As mentioned earlier, we shall consider (a) an infinite square array and (b) an isolated cluster of randomly placed
emitters.

\subsection{Infinite square array}

Consider an infinite square array with lattice constant $c$ and an emitter of height $h = 1500\mu$m and apex
radius of curvature $R_a = 1.5\mu$m. The lattice constant $c$ is now varied and the apex field enhancement
factor $\gamma_a$ is calculated using (i) the LCM prediction of Eq.~(\ref{eq:gamN0}) with $R = 100h$  and (ii) COMSOL with Neumann boundary
condition at the anode with the anode-cathode separation fixed at $D = 5h$. The relative error

\be
   \text{Error} (\%) = \frac{|\gamma_a^{LCM} - \gamma_a^{COMSOL}|}{\gamma_a^{COMSOL}} \times 100
\ee

\begin{figure}[htb]

\hspace*{-1.0cm}\includegraphics[width=0.55\textwidth]{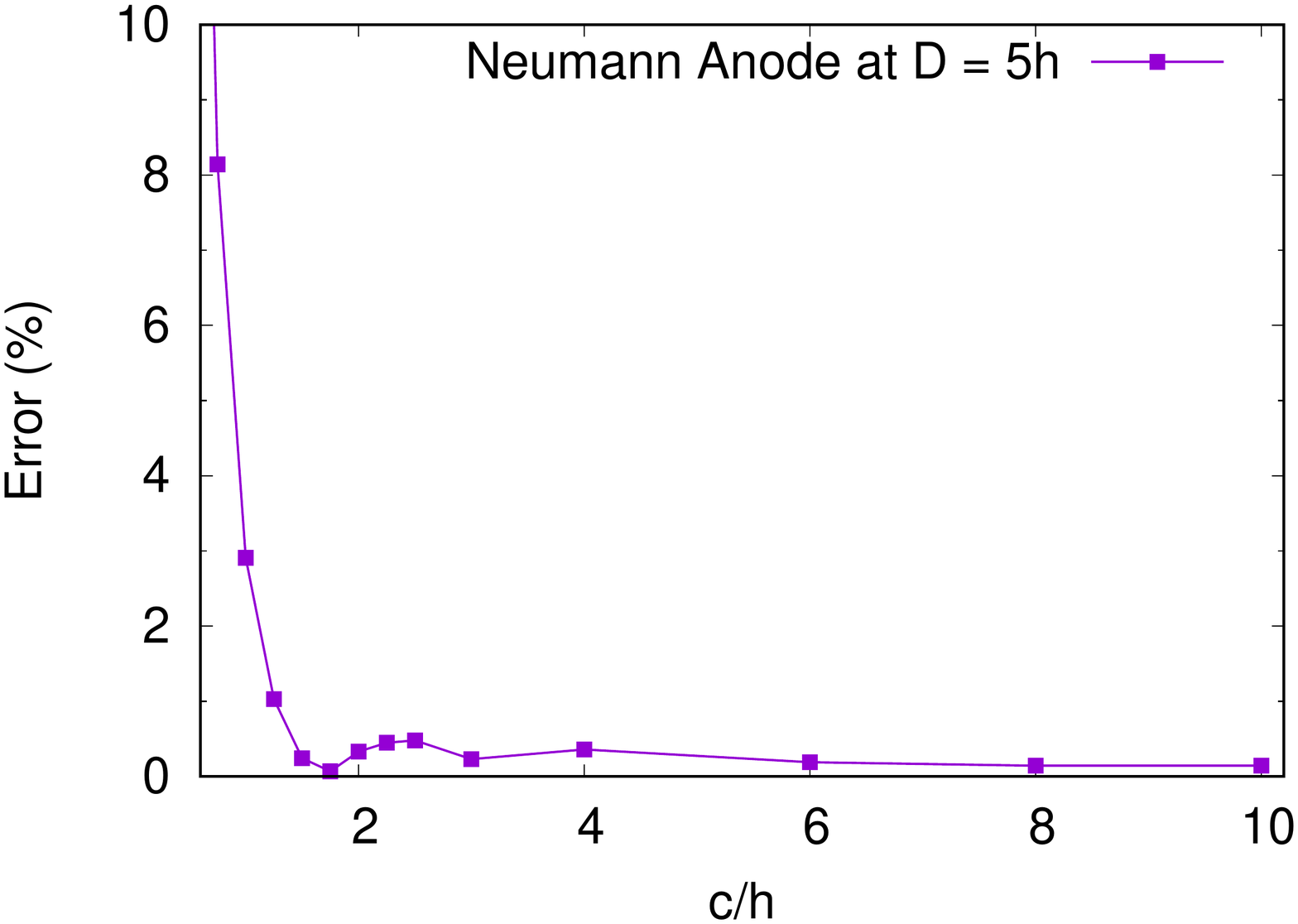}

\caption{The relative error in apex enhancement factor evaluated using the LCM prediction of Eq.~(\ref{eq:gamN0}). The `exact' result is calculated using COMSOL. The emitter height $h = 1500\mu$m while $R_a = 1.5\mu$m.
  }
  \label{fig:error_lattice}
\end{figure}

\noi
is calculated at different values of lattice constant $c$. The result is shown in Fig.~\ref{fig:error_lattice}.

The error is about $2.9\%$ for $c = h$ and about $8.1\%$ for $ c = 0.75h$. It falls to about $1\%$ for $c=1.25h$
and remains less than $0.5\%$ for $c \ge 1.5h$. For larger values of $c$, the error becomes smaller than 0.1\%.
Note that in the region where the optimal current density is expected to lie, the error remains small.
Thus, the analytical result (Eq.~(\ref{eq:gamN0})) based on the line charge model can be used to calculate
the net emitted current accurately for a large array, provided the density of emitters is such that $c \ge h$.

\subsection{Cluster of random emitters}

Randomly placed emitters pose a greater challenge insofar as verification of Eq.~(\ref{eq:gamN0}) is concerned.
It is difficult to model these using a finite element software such as COMSOL since, even for reasonable computational
resources, the number of emitters may be limited to about 25-30 depending on the $h/R_a$ ratio. Note that
the AFEF calculation for an array or cluster requires 3-dimensional modeling and the demands on resources increases
as the curvature at the apex increases. Thus, an analytical model, if validated and found to be reasonably accurate,
can serve as a useful tool in optimizing the emitter density of a LAFE.

In the present context, an isolated cluster of randomly placed emitters is sought to be modelled. The computational
boundary must therefore be chosen to be sufficiently far away if a standard Neumann boundary is to be used
in the $X,Y$ directions. Typically, the domain considered is $[-10h,10h]$ in the $X,Y$ directions and
$[0,5h]$ in the $Z$ direction. The emitters are limited to a patch  at the centre in the $X,Y$ plane
such that the mean spacing equals $h$. We have considered 2 such realizations, one having 5 emitters and the other
having 10. We present here the results for 10 emitters.

\begin{figure}[htb]
\vskip -0.35 in
\hspace*{-1.0cm}\includegraphics[width=0.55\textwidth]{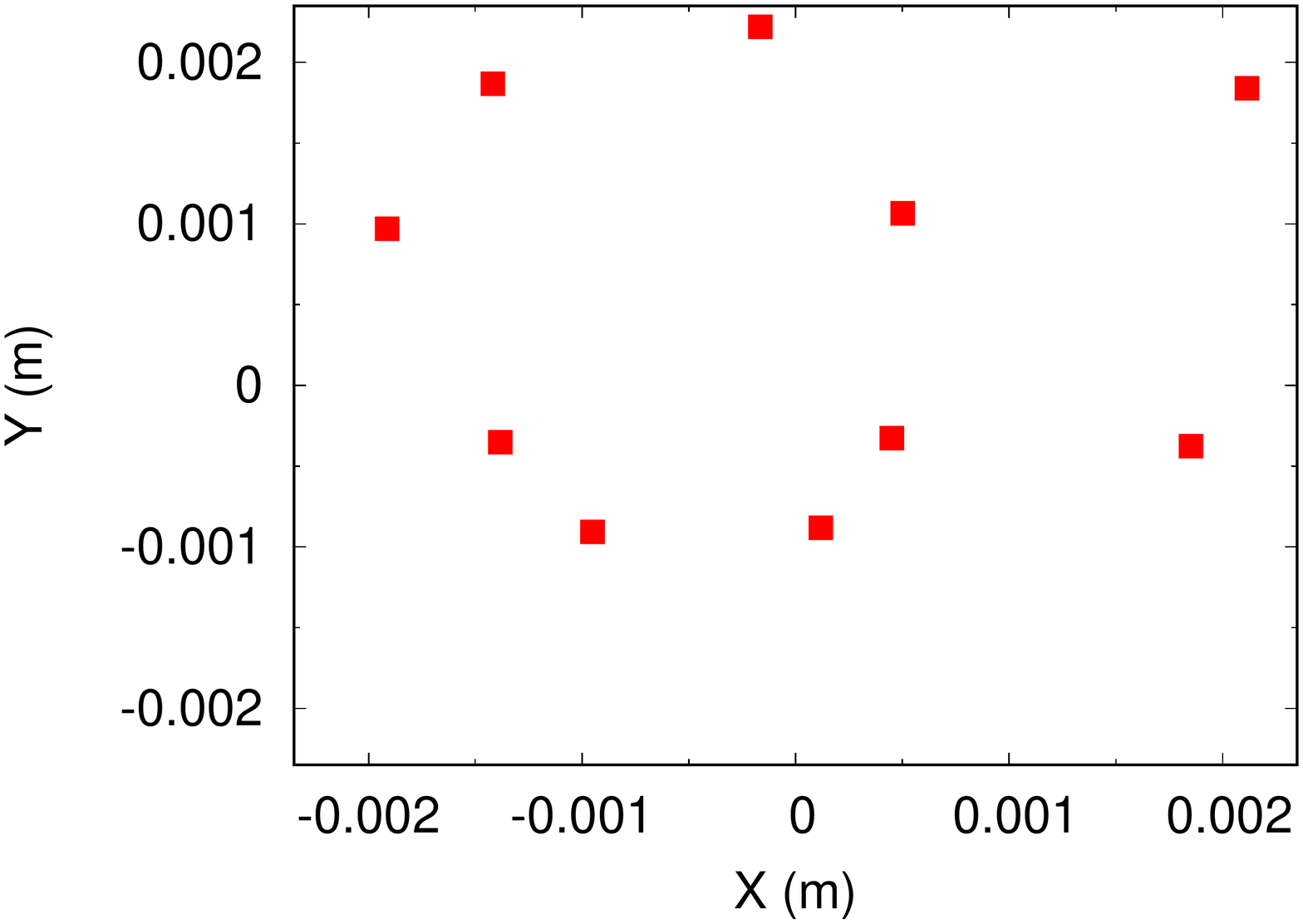}
\vskip -0.3 in
\caption{A cluster of 10 randomly placed emitters in the $XY$ plane}
  \label{fig:rand10}
\end{figure}

The distribution of emitters in the $XY$ plane  is shown in Fig.~\ref{fig:rand10}. The apex field enhancement
factor calculated using COMSOL and the line charge model is shown in Table~\ref{tab:gam} along with the error.
Clearly, the error is small when the emitter has larger AFEF value (less shielding). This suggests that the
error in net emission current may be smaller for a random cluster than a regular infinite array with the same
mean spacing.

\begin{table}[htb]
 \begin{center}
   \caption{A comparison of apex field enhancement factors using COMSOL and LCM at different emitter locations in a cluster.}
   \vskip 0.25 in
    \label{tab:gam}
    \begin{tabular}{|c|c|c|c|c|}
      \hline
      X (m) & Y (m) & $\gamma_a^{COMSOL}$ & $\gamma_a^{LCM} $& Error (\%) \\ \hline \hline
      $-1.418\times 10^{-3}$  & $1.867\times 10^{-3}$ & 274.97 & 271.81 & 1.15 \\ \hline      
      $1.186\times 10^{-4}$ & $-8.817\times 10^{-4}$ & 258.35 & 250.64 & 2.98 \\ \hline
      $-9.511\times 10^{-4}$ &$-9.063\times 10^{-4}$ & 262.71 & 255.78 & 2.63 \\ \hline
      $-1.913\times 10^{-3}$ &$9.696\times 10^{-4}$ & 275.49 &272.08 &  1.23 \\ \hline
      $2.114\times 10^{-3}$ & $1.839\times 10^{-3}$ & 294.34 & 293.54 & 0.275 \\ \hline
      $-1.384\times 10^{-3}$ & $-3.521\times 10^{-4}$ &264.13 & 257.51 & 2.50 \\ \hline
      $4.507\times 10^{-4}$ & $-3.264\times 10^{-4}$ & 254.63 & 246.74 & 3.10 \\ \hline
      $-1.652\times 10^{-4}$ & $2.220\times 10^{-3}$ & 278.38 & 275.79 & 0.932 \\ \hline
      $1.852\times 10^{-3}$ & $-3.759\times 10^{-4}$ & 285.85 & 283.91 & 0.677 \\ \hline
      $5.022\times 10^{-4}$ & $1.065\times 10^{-3}$ & 268.47 & 263.78 & 1.75 \\ \hline
  \end{tabular}
\end{center}
\end{table}

The net field emission current can be calculated using the apex field enhancement factor and the generalized
cosine law of local field variation\cite{db_ultram,db_physicaE} around the emitter
apex as shown in [\onlinecite{db_dist}] for emitters
having $R_a > 100$nm. For the random cluster under consideration, the error in net emission current density 
is a nominal $15.5\%$ at a macroscopic field of 17.5 MV/m while for the infinite array having $c = h$,
the roughly $3\%$ error in the LCM-AFEF shifts the current density by a factor of 2. Both results
are acceptable (given the inherent uncertainties in field emission theory predictions) though the random
distribution has an edge insofar as the current estimation is concerned. These errors are likely to decrease
as the mean separation or lattice constant increases and also with an increase in macroscopic field strengths.

\section{Summary and Conclusions}

The analytical predictions of the line charge  model for large area field emitters has been the subject of
investigation in the paper. The model for LAFE uses hemi-ellipsoid emitters
as the basic building blocks. It allows a computation of the apex field enhancement factor (AFEF) for any emitter in terms of
normalized pair-wise distance to the emitters in its LAFE neighbourhood. As a purely geometric
model which ignores charge distribution details,
the approximate values of AFEF that it provides has been the subject of scrutiny in this study. Our results
show that if emitters are separated by average distances approximately equal to or greater than the height of the emitters, the
errors in the prediction of LCM model are small and the values of net emitted current are acceptable.
In the process of establishing this, we also demonstrated the domain size necessary for simulating a large area
field emitter, both from the point of view of finite element methods and the line charge model.

The results are particularly encouraging for field emission since the line charge model performs well for
emitters that contribute significantly to the net current density. It can thus be used to study large clusters of emitters
which are otherwise inaccessible to computations due to the resources involved.

Practical emitter shapes used in field emission may vary from cylindrical structures to cones with the added possibility of
differently shaped endcaps. The hemi-ellipsoid can be used to approximate these keeping the apex radius of
curvature invariant. This will undoubtedly lead to errors in the AFEF calculation of single emitters since the
emitter base plays an important role. For a random LAFE however, it is worth investigating if such finer
points take a back seat.

%\section{Acknowledgements}

\vskip 0.05 in
%$\;$\\
\section{References} 
\vskip -0.25 in
%\begin{references}


\begin{thebibliography}{99}
\bibitem{spindt68} C.~A.~Spindt, J.~Appl.~Phys., 39, 3504 (1968).
\bibitem{spindt76} C.~A.~Spindt, I.~Brodie, L.~Humphrey, and E.~R.~Westerberg, J.~Appl.~Phys. 47, 5248 (1976).
\bibitem{parmee} R.~J.~Parmee, C.~M.~Collins, W.~I.~Milne, and M.~T.~Cole, Nano Convergence 2, 1 (2015).
\bibitem{basu2015} A. Basu, M. E. Swanwick, A. A. Fomani, and L. F. Velásquez-García, J. Phys. D Appl. Phys. 48, 225501 (2015). 
\bibitem{cole2016} M. T. Cole, R. J. Parmee, and W. I. Milne, Nanotechnology 27, 082501 (2016).
\bibitem{whaley2018} D.~R.~Whaley, C.~M.~Armstrong, C.~E.~Holland, C.~A.~Spindt, P.~R.~Schwoebel, 31st International Vacuum Nanoelectronics Conference (IVNC) (2018), 10.1109/IVNC.2018.8520271. 
\bibitem{FN} R.~H.~Fowler and L.~Nordheim, Proc. R. Soc. A 119, 173 (1928).
\bibitem{murphy} E.~L.~Murphy and R.~H.~Good, Phys. Rev. 102, 1464 (1956).
\bibitem{forbes} R.~G.~Forbes, App. Phys. Lett. 89, 113122 (2006).
\bibitem{jensen_ency} K.~L.~Jensen, {\it Field emission - fundamental theory to usage},
  Wiley Encycl. Electr. Electron. Eng. (2014).
\bibitem{forbes2003} R.~G.~Forbes, C.J.~Edgcombe and U.~Valdr\`{e}, Ultramicroscopy 95, 57 (2003).
\bibitem{db_fef} D.~Biswas, Phys. Plasmas 25, 043113 (2018).
\bibitem{read_bowring} F.~H.~Read and N.~J.~Bowring, Nucl. Instrum. Meth. Phys. Res. A 519, 305 (2004).
\bibitem{harris15} J.~R.~Harris, K.~L.~Jensen, D.~A.~Shiffler and J.~J.~Petillo, Appl. Phys. Lettrs. 106, 201603 (2015).
\bibitem{jap2016} D.~Biswas, G.~Singh and R.~Kumar, J.~App.~Phys. 120, 124307 (2016). 
\bibitem{db_rudra} D.~Biswas and R.~Rudra, Phys. Plasmas, 25, 083105 (2018).
\bibitem{kosmahl} H.~G.~Kosmahl, IEEE Trans. Electron Devices 38, 1534 (1991).
\bibitem{pogorelov} E.~G.~Pogorelov, A.~I.~Zhbanov, and Y.-C.~Chang, Ultramicroscopy 109, 373 (2009).
\bibitem{forbes_2018a} R.~G.~Forbes, J.~Appl.~Phys. 120, 054302 (2016).
\bibitem{agnol} T. A. de Assis and F. F. Dall'Agnol, J.~Vac.~Sci.~Tech. B, 37, 022902 (2019).
\bibitem{db_ultram} D.~Biswas, G.~Singh, S.~G.~Sarkar and R.~Kumar, Ultramicroscopy 185, 1 (2018).
\bibitem{db_physicaE} D.~Biswas, G.~Singh, R.~Ramachandran, Physica E 109, 179 (2019).
\bibitem{db_dist} D.~Biswas, Phys. Plasmas 25, 043105 (2018).
  
\end{thebibliography}
\end{document}